# Mid-Infrared spectroscopy of impactites from the Nördlinger Ries impact crater

**Corresponding Author:** Andreas Morlok, Institut für Planetologie, Wilhelm-Klemm-Str. 10, 48149 Münster, Germany. Email: morlokan@uni-muenster.de, Tel. +49-251-83-39069

Aleksandra Stojic, Institut für Planetologie, Wilhelm-Klemm-Str. 10, 48149 Münster, Germany. Email: a.stojic@uni-muenster.de;

Isabelle Dittmar, Hochschule Emden/Leer, Constantiaplatz 4, 26723 Emden, Germany, Email: isabelle.dittmar@hs-emden-leer.de

Harald Hiesinger, Institut für Planetologie, Wilhelm-Klemm-Str. 10, 48149 Münster, Germany. Email: hiesinger@uni-muenster.de

Manuel Ahmedi, Institut für Planetologie, Wilhelm-Klemm-Str. 10, 48149 Münster, Germany. Email: ahmedi@uni-muenster.de

Martin Sohn, Hochschule Emden/Leer, Constantiaplatz 4, 26723 Emden, Germany, Email: martin.sohn@hs-emden-leer.de

Iris Weber, Institut für Planetologie, Wilhelm-Klemm-Str. 10, 48149 Münster, Germany. Email: sonderm@uni-muenster.de

Joern Helbert, Institute for Planetary Research, DLR, Rutherfordstrasse 2, 12489 Berlin, Germany, Email: joern.helbert@dlr.de

# ABSTRACT

This study is part of an effort to build a mid-infrared database (7-14µm) of spectra for MERTIS (Mercury Radiometer and Thermal Infrared Spectrometer), an instrument onboard of the ESA/JAXA BepiColombo space probe to be launched to Mercury in 2017.

Mercury was exposed to abundant impacts throughout its history. This study of terrestrial impactites can provide estimates of the effects of shock metamorphism on the mid-infrared spectral properties of planetary materials.

In this study, we focus on the Nördlinger Ries crater in Southern Germany, a well preserved and easily accessible impact crater with abundant suevite impactites. Suevite and melt glass bulk samples from Otting and Aumühle, as well as red suevite from Polsingen were characterized and their reflectance spectra in mid-infrared range obtained. In addition, in-situ mid-infrared spectra were made from glasses and matrix areas in thin sections. The results show similar, but distinguishable spectra for both bulk suevite and melt glass samples, as well as in-situ measurements.

Impact melt glass from Aumühle and Otting have spectra dominated by a Reststrahlen band at 9.3-9.6 µm. Bulk melt rock from Polsingen and bulk suevite and fine-grained matrix have their strongest band between 9.4 to 9.6 µm. There are also features between 8.5 and 9 µm, and 12.5 - 12.8 µm associated with crystalline phases. There is evidence of weathering products in the fine-grained matrix, such as smectites. Mercury endured many impacts with impactors of all sizes over its history. So spectral characteristics observed for impactites formed only in a single impact like in the Ries impact event can be expected to be very common on planetary bodies exposed to many more impacts in their past. We conclude that in mid-infrared remote sensing data the surface of Mercury can be expected to be dominated by features of amorphous materials.

## 1. Introduction

The aim of this study is to provide mid-infrared reflectance spectra with special emphasis on the region from 7μm – 14μm for a range of impactites for the application in planetary remote sensing. We generate these spectra for a database for the ESA/JAXA BepiColombo mission to Mercury (Benkhoff et al., 2010). Onboard is a mid-infrared spectrometer (MERTIS-Mercury Radiometer and Thermal Infrared Spectrometer). This unique device allows mapping spectral features in the 7-14 μm range, with a spatial resolution of ~500 m (Helbert et al., 2009; Benkhoff et al., 2010; Hiesinger et al., 2010).

Infrared spectroscopy provides a means to characterize the mineralogy of rocks via remote sensing, in contrast to Gamma-Ray, Neutron and X-ray spectrometers, which determine elemental compositions (Pieters and Englert, 1993). Thus, IR spectroscopy provides a central analytical tool to determine the mineralogy of remote planetary surfaces. In order to correctly interpret the remote sensing data, laboratory spectra of natural, synthetic rocks and minerals have to be collected to compare them to the spectra that will be obtained from MERTIS, once BepiColombo enters the Hermean orbit (Maturilli et al., 2008; Hiesinger et al., 2010). Comparable earlier instruments were the Thermal Emission Spectrometer (TES) on the Mars Global Surveyor (Christensen et al., 2005) and the Thermal Emission Imaging System (THEMIS) on the Mars Odyssey orbiter (Christensen et al., 2004). The lunar surface was mapped in the mid-infrared with the DIVINER Lunar Radiometer Instrument on the Lunar Reconnaissance Orbiter (Paige et al., 2010). The OSIRIS-REx Thermal Emission Spectrometer (OTES) will map asteroid Bennu (1999 RQ36) (Hamilton and Christensen, 2014), and the Thermal Infrared Imager (TIR) onboard Hayabusa 2 will map asteroid 1999JU3 (Okada et al., 2015).

The Ries crater in southern Germany (Fig.1) provides well preserved layers of impactites and is one of the best studied impact sites in the world (von Engelhardt, 1995). The 14.6 Myr old crater with a diameter of 24 kilometers (Buchner et al., 2010) offers the whole range of impact-associated rocks and minerals (von Engelhardt, 1990). Large impact events play a key role in surface modification processes in effect on most terrestrial planets and their moons (Hörz and Cintala, 1997). Thus, the investigation of terrestrial impact processed rocks and understanding how these processes affect the spectral properties of the resulting impact generated rocks and melt glass is important for the interpretation of infrared data from the surfaces of other planetary bodies. Given the characteristics of surface regolith, we need spectral data from different grain size fractions. This is necessary because grain size variations affect the corresponding spectra immensely by reducing the spectral contrast of features and creating additional features at small grain sizes (e.g., Salisbury and Eastes, 1985; Salisbury and Wald, 1992; Mustard and Hayes, 1997; Ruff and Christensen, 2002).

Reflectance and emission studies about spectral features in the mid-infrared in minerals undergoing high shock metamorphism including formation of melt glass were made on experimentally shocked samples, e.g., on anorthosite, pyroxenite, basalt, and feldspar (Johnson et al. , 2002, 2003, 2007, 2012; and Jaret et al. ,2015). In these studies, especially feldspar-rich material showed loss and degradation of features, as well as band shifts with increasing structural disorder resulting from increasing shock pressure. Moroz et al. (2010) analyzed impact glasses from laser pulse experiments with Martian soil analogue JSC Mars-1. Similar, Basilevsky et al. (2000) and Morris et al. (2000) spectrally studied melt glass also made from Martian soil analogs. Byrnes et al. (2007), and Lee et al.,(2010) measured synthetic quartzofeldspathic glasses and Dufresne et al. (2009), Minitti et al. (2002), Minitti and Hamilton (2010) measured synthetic glass with basaltic to intermediate composition. In general, the resulting mid-infrared reflectance or emission spectra display a dominant feature in the 9.2-10.5 µm range. Pollack et al. (1973), Crisp et al., 1990; Nash and Salisbury (1991) and Wyatt et al. (2001), studied obsidian or basalt. Further infrared reflectance and emission spectra of natural impact glass and tektites were made, e.g., by Thomson and Schultz (2002), Gucsik et al. (2004), Faulques et al. (2005) and Palomba et al. (2006). Wright et al. (2011), Basavaiah and Chavan (2013), and Jaret et al. (2013) investigated shocked bulk material from the Lonar Crater, India, showing mainly a simple spectrum with a dominating feature in the ~9.4 - 10.2 µm range.

Complementary infrared transmission and absorbance spectra of related materials were also made, e.g, of experimentally shocked feldspar materials (Stöffler and Hornemann, 1972; Ostertag, 1983) and of CM2 chondrite Murchison (Morlok et al., 2010). Glasses with feldspathic composition produced in static high pressure experiments were measured by, e.g., Iiishi et al. (1971), Velde et al. (1987), and Williams and Jeanloz (1988, 1989). In addition, transmission infrared analyses of natural impact glass and tektites for studying water contents were made by Beran and Koeberl (1997), for identification purposes (Fröhlich et al., 2013) or for astrophysical studies of circumstellar dust (Morlok et al., 2014). King et al. (2004) provided an overview of silicate glass analysis with various mid-infrared techniques.

Furthermore, there are also numerous studies how impact shock affects the spectral properties in the visible and near-infrared, such as Johnson and Hörz (2003), Adams et al. (1979), and Bruckenthal and Pieters (1984) for experimentally shocked feldspars or enstatite. Moroz et al. (2009) studied synthetic glasses with Martian soil composition as impact melt analogs, and Cannon and Mustard (2015) identified glass-rich impactites on Mars. Bell et al. (1976) and Stockstill-Cahill et al. (2014) analyzed lunar glass analogues, while Keppler (1992) studied synthetic silicate glasses with albite and diopside composition. Schulz and Mustard (2004) studied terrestrial impact melt rocks.

Shock metamorphism begins with fracturing and brecciation of the rocks at lower pressures; above 2 GPa the mineral phases start to change. Shatter cones and conical fracturing patterns form. Between 8 and 25 GPa, planar deformation features appear along with microscopic changes in the crystal structure due to impact shock in, e.g., quartz and feldspar. At pressures >25 GPa, shock metamorphism continues with the solid-state transformation of minerals into diaplectic glasses, which are amorphous materials generated at pressures up to 40 GPa. At pressures over 35 GPa, partial melting of phases begins and at over 60 GPa rocks completely melt, followed by vaporization at pressures above 100 GPa (e.g., Stöffler, 1966, 1971, 1984; Chao, 1967; von Engelhardt and Stöffler, 1968; Stöffler and Langenhorst, 1994; French, 1998). Also, high pressure mineral polymorphs form; stishovite and coesite form from quartz at pressures from 12-15 GPa and over 30 GPa, respectively. Carbon is transformed into diamond at 13 GPa (Stöffler and Langenhorst, 1994; French, 1998).

In the impactite suevite, shock metamorphism can be divided into stages 0 – IV.  Stage 0 ranges from 0-10 GPa, stage I (10-35 GPa) and II (35-60 GPa) cover changes in crystal structure and partial melting; stage IV pressures over 60 GPa result in melt glasses (Stöffler, 1971 and 1984; French, 1998; Stöffler and Grieve, 2007; Stöffler et al., 2013). Suevite consists of materials of all five shock stages (Stöffler et al., 2013). Thus the infrared spectra obtained from the suevites can be expected to contain a mixture of spectral features depending on the level of shock metamorphism the material underwent.

In this study, we focus on impactites from the Ries crater, since the crater site allows easy access to a range of naturally shocked rocks and minerals (von Engelhardt, 1990, 1995). Prior to the impact event, the Ries area was covered with sediments mainly consisting of limestones. The underlying crystalline basement is dominated by granites and gneisses, with significant amounts of amphibolite (von Engelhardt and Graup, 1984; von Engelhardt, 1997). The average chemical composition of very homogeneous impact melts (shock stage IV; Stöffler et al, 2013) is similar to the composition of the modelled crystalline basement rock clasts based on the fallout suevite, identifying these rocks as the main source of the homogenized melt (Staehle, 1972; von Engelhardt et al., 1984 , 1995, 1997; Vennemann et al., 2001).

We focus on the suevite, the top layer of ejecta in the Ries that was deposited on all other impact ejecta, Bunte Breccia (e.g. Abadian, 1972; Hörz et al., 1983), megablocks (e.g. Sturm et al., 2015) and polymict crystalline breccia (e.g. von Engelhardt, 1997). The most voluminous ejecta is the Bunte Breccia, consisting mainly of sedimentary material. Megablocks are rocks in the size range from 25 m to kilometers, consisting of sedimentary (limestone) and crystalline materials. Both Bunte Breccia and megablocks show low degrees of shock metamorphism. Polymict crystalline breccias are mixtures of basement rocks with shock stages up to stage II. Suevite is the only ejecta layer to contain material in all stages of shock metamorphism (e.g., von Engelhardt, 1969, 1990, 1997). Moldavites, a type of tektites formed in the Ries impact are an additional type of ejecta that was deposited 350 km east of the impact site (e.g., von Engelhardt, 1987).

The mostly granitic and felsic petrology of the crystalline basement of the Ries, which controlled the composition of the suevite (von Engelhardt et al., 1990, 1997), is quite different compared to that of Mercury and other planetary surfaces in the Solar System. Also, granite in general is very rare outside Earth (Bonin, 2012). Based on MESSENGER data, the best terrestrial analogues for the surface of Mercury are basalts or ultramafic komatiites (Nittler et al., 2011; Stockstill-Cahill et al., 2012; Charlier et al., 2013; Maturilli et al., 2014), but natural shocked forms are rare on Earth (e.g. Wright et al., 2011). However, a spectral study of naturally shocked granitic material is of interest for our remote sensing purposes because it gives insight into the spectral characteristics of impactites and their components, while naturally shocked basalts are rare (Wright et al., 2011). Furthermore, the easy access to the Ries site gives access to larger, representative amounts of sample material. Although rare, granitic materials have been recognized on other planetary bodies. Granitic material has been found as fragments and clasts in lunar samples (e.g., Warren et al., 1983; Jolliff et al., 1999; Shervais and McGee, 1999; Seddio et al., 2015). Granitoid or felsic materials may also occur on the mostly basaltic surface of Mars (e.g., Christensen et al., 2005; Bandfield, 2006, Ehlmann and Edwards, 2014; Sautter et al., 2014, 2015). There are also indications for felsic, granitoid material on Venus (Müller et al., 2008; Gilmore, 2015).

Suevites are divided into crater or fallback suevite, occurring inside the inner ring of the Ries, and outer suevite, encompassing suevite found outside of the inner part of the crater (Pohl et al., 1977, Stöffler, 1977, 2013). Normal suevite in this study is outer suevite that originates from the Otting and Aumühle quarries (Fig.1, 2a,b). It mainly consists of three components: a porous matrix of fine-grained rock, melt glass, and crystalline basement rocks displaying all stages of shock metamorphism ( Stöffler, 1966; von Engelhardt and Graup, 1984; Stöffler et al., 2013). Red suevite (Fig. 2c) from the Polsingen quarry is a rare variation of the outer suevite. Its red color is a characteristic owing to high hematite content (Stöffler et al., 2013). In the red suevite, the actual groundmass consists of impact melt, in which fragments are embedded (Reimold et al., 2010). Also, chemical differences between normal suevite and red suevite ($Na_2O$, $K_2O$) were reported by Reimold et al. (2013) and Stöffler (2013). In addition to the bulk materials, we are also interested in the components of the suevite. In particular, we studied the melt glasses (Fig.2a, b), i.e., completely shock-melted and quenched rock material. Furthermore, we looked at the fine-grained matrix itself, which was affected by weathering processes (von Engelhardt, 1995; Stöffler et al., 2013).

## 2. Samples and Techniques

### 2.1 Sample Selection

We selected three bulk suevite samples from three different Ries localities of the suevite layer (von Engelhardt and Stöffler, 1974; von Engelhardt et al., 1995, 1997; Bayerisches Geologisches Landesamt, 2004): normal crater suevite from the Otting (samples Otting Bulk 1-3) and Aumühle (Aumühle Bulk4, Bulk 18, Bulk 18 Matrix) quarries, and red suevite (samples Polsingen 2-4) from a small outcrop in Polsingen (outer suevite) (von Engelhardt, 1997; Reimold et al., 2010). In addition, we used material from three separated impact melt glasses: from a large, 30 cm sized glass 'bomb' (Otting Glasbombe) and a separated glass 'Flädle' from Otting (Otting Glas 1), and another glass separated from the Aumühle suevite (Aumühle 13).

In addition, for quantitative chemical and additional IR-microscopy analyses, polished thin sections of Aumühle, Otting and Polsingen were produced from the bulk sample rocks. The vitreous state of the glass was confirmed using polarized light microscopy by the total extinction of the glass under crossed polarizers (Fig.2a-c). However, the two Otting samples show brownish halos or 'Schlieren', signs of incipient devitrification by the formation of small crystallites of pyroxenes, feldspar, oxides and possibly incipient signs of alteration phases like clays (see discussion)(von Engelhardt et al., 1995).

**2.2 Sample Preparation**

We used aliquots of larger amounts of material to avoid bias due to larger fragments/components thus ensuring a homogeneous and representative sample. The original sample masses were greater than 100 grams for each sample. Bulk samples were first powdered in steel and agate mortars. Subsequently, the powders were cleaned in acetone and dry sieved into four size fractions: 0-25 µm, 25-63 µm, 63-125 µm and 125-250 µm. This was done using an automatic Retsch Tap Sieve; each size fraction was dry sieved for at least one hour. In order to remove clinging fines, the larger two fractions were cleaned with acetone.

For additional in situ measurements involving optical microscopy, micro-FTIR, and Scanning Electron Microscopy (SEM) measurements, we used thin sections of representative blocks of the samples. The thin sections were polished to ~30 µm using standard procedures for petrological thin sections, ensuring a high specular reflectance from a flat surface from which especially microscopic IR investigations benefit greatly.

**2.3 Optical Microscopy**

Overview images of the samples in normal light and between crossed polarizers were obtained with the KEYENCE Digital Microscope VHX-500F. Light microscopy allows rapid assessment of the general homogeneity, as well as first mineral identification in the samples. Images were made using a magnification of 10.

**2.4 Infrared Analyses**

**2.4.1 Diffuse Reflectance Powder Analyses**

To ensure diffuse reflectance, the size fractions were gently placed in aluminum sample cups (1 cm diameter), and the surface flattened with a spatula following a similar procedure described by Mustard and Hayes (1997). For mid-infrared analyses from 2-20 µm, we used a Bruker Vertex 70 infrared system with a MCT detector at the Infrared and Raman for Interplanetary Spectroscopy (IRIS) laboratory at the Institut für Planetologie in Münster. To avoid sample surface disturbance due to pore collapse during evacuation, we avoided analyses under near vacuum and measured under normal atmosphere. As a consequence, water and atmosphere related features slightly affect the spectral range of interest near 7 µm. To ensure a high signal-to-noise ratio, we accumulated 512 scans for each size fraction. For background calibration a diffuse gold standard was applied. For the MERTIS database, we obtained analyses in a variable geometry stage (Bruker A513) in order to emulate various observational geometries of the orbiter. The data presented here were obtained at 30° incidence (i) and 30° emergence angle (e).

The spectra of this study are intended to be compared with remote sensing data in the thermal infrared. Emission and reflectance spectra can be compared using Kirchhoff's law: $\varepsilon = 1 - R$ (R=Reflectance, $\varepsilon$ = Emission) (Nicodemus, 1965). This relation works very well for the comparison of directional emissivity and directional hemispherical reflectance (Hapke, 1993, Salisbury et al., 1994)**.**

However, in order to directly relate directional emissivity with reflectance by using Kirchhoff's law, the reflected light in all directions has to be collected (Thomson and Salisbury, 1993). A bi-directional, variable mirror set-up was used for this study without a hemisphere integrating all radiation. This has to be kept in mind when comparing the results in a quantitative manner with emission data (Salisbury et al., 1991; Christensen et al., 2000). In the case of the spectra obtained with the micro-FTIR in specular reflectance mode, the diffusely scattered part of the light is very small, since highly polished glass was analyzed. Although specular reflectance is not entirely relatable to the directed emission, as required by Kirchhoff's law, the differences will be mainly in spectral contrast, otherwise (e.g. band shape and positions) the data will be comparable (Ramsey and Fink, 1999; Byrnes et al, 2007; Lee et al., 2010).

Although the spectral range of interest for the database is from 7-14 µm, we measured our powder from 7-20 µm (Fig. 4a-c), since features of interest can appear at longer wavelengths. The signal of the detector used becomes weak at wavelengths above 18 µm, resulting in a low signal to noise ratio. Above 19 µm, the spectral features are mainly random noise. The analyses were conducted under ambient pressure, which possibly affected the water bands at ~3 and ~6 µm. We therefore only present representative features from 2-7 µm for each sample in Fig.4d. The impactites analyzed in this study are often mixtures of various mineral phases. So in order to identify specific mineral bands in our laboratory spectra, we used spectra from the Arizona State University Thermal Emission laboratory (Christensen et al., 2000) and the Johns Hopkins ASTER laboratory (Baldridge et al. 2009).

### 2.4.2 In-situ FTIR Microspectroscopy

For in situ analyses, we used a Bruker Hyperion 2000 IR microscope attached to the external port of a Bruker Vertex 70v at the Hochschule Emden/Leer. Here a 1000×1000 µm$^2$ sized aperture was applied to obtain analyses of interesting features with in situ reflectance spectroscopy on polished thin sections. For each spectrum, 128 scans were added. A gold mirror was used for background calibration. The analyses were made in the range from 2-15 µm. Since all features below 7 µm are very weak, we present results in the range of interest, 7-14 µm (Fig.5).

The small shift between powder and microscope analyses observed (see 3. Results) is probably due to the inherent differences among the samples (polished thin sections compared to powders) and the different optical set-ups of the techniques. Differences in the positions between the CF in powdered and solid samples were also observed by Cooper et al. (2002).

**2.5 SEM EDX Analyses**

In order to document the suevites and their components, we used a JEOL 6610-LV Scanning electron microscope equipped with a silicon drift Oxford EDX (Energy Dispersive X-Ray Spectroscopy) system to obtain micrographs of particular areas of interest and perform quantitative chemical analyses. Each chemical analysis was quantified with an ASTIMEX™ standard set for major elements prior to the measurement. Beam current stability was controlled and measured before each analysis using a Faraday cup. The calibration was confirmed by re-analyzing the standards after the calibration procedure. For analyses of the chemical composition of melt glass and fine-grained matrix areas, we analyzed areas of 100 x 100 µm$^2$ using 90 seconds integration times. The rather small area was chosen to maintain comparability between amorphous parts and fine-grained matrix areas. Otherwise it would be difficult to obtain representative measurements of larger areas due to abundant cracks, gaps, holes, and veins, which are often re-filled by secondary alteration phases. A broad beam and shorter integration times are helpful to measure volatile elements correctly.

# 3. Results

## 3.1 Optical Images

Optical images (Fig2a-c) give an overview of the thin sections for the three investigated samples. The Otting (Fig.2a) and Aumühle (Fig.2b) samples in the transmitted light show the components of the normal suevite, the darker melt glasses, embedded in a brighter matrix, which consists of fine fragments of glass, rock and also secondary alteration products. Under crossed polarizers, the glasses are recognizable by their nearly black appearance due to complete extinction, while the fine mixture of the matrix appears brighter. The Polsingen sample (Fig.2c), being a coherent melt rock, consists entirely of a red groundmass with abundant larger and smaller fragments in transmitted light. Brownish rims around minerals and holes indicate significant weathering. In the image under crossed polarizers, the groundmass appears dark, a few crystalline fragments are visible by their brighter appearance.

## 3.2. SEM/EDX Analyses

Melt glass and fine-grained matrix area measurements in the suevites are shown in Table 1. Due to porosity and volatile contents (water) the total analysis of matrix material in weight (wt)% is lower than 100 wt%. The melt glasses also show lower totals, which is due to crystallized or oxidized inclusions as well as inevitable cracks and veins.

For a better comparison with earlier data, we plot characteristic oxide ratios for MgO, FeO, CaO, $Na_2O$, $K_2O$ of our SEM/EDX results with those from the literature (Fig. 3) (Stöffler et al., 2013). Results for melt glass in Aumühle and Otting are very similar to each other and to earlier studies of glass from outer suevite. The results for the fine-grained matrix analyses cluster in an area overlapping with the composition of outer suevite (Fig. 3) (Stöffler et al., 2013). The analyses plotting outside the area are explained by the lack of larger rock fragments within the analyzed areas. This resulted in a higher content of fine grained secondary phases typical for the matrix, such as clay minerals, which is also indicated by elevated $Al_2O_3$ contents (Tab. 1) (Stöffler et al., 2013). Otherwise, the compositions of the fine-grained matrices can be explained as a mixture of the endmember glass and various basement rock fragments. Melt rock from the red suevite in Polsingen plots slightly below the studies for bulk red suevite (Stöffler et al., 2013). This is probably due to the high degree of weathering of the samples from the only accessible outcrops, resulting in increased alkali ($Na_2O$, $K_2O$, $CaO$) concentrations (Reimold et al., 2013; Stöffler, 2013).

### 3.3 Mid-Infrared Analyses of Powders

The powder spectra of the bulk normal suevite from Otting and Aumühle (Fig.4a, Tab. 2a) show very similar band shapes and intensities for all samples, reflecting the high chemical homogeneity already indicated by SEM/EDX. The Christiansen Features (CF) of the bulk Otting and Aumühle bulk suevites are between 7.4 and 8.1 µm, respectively, for all size fractions within one sample group. In the finest fraction from 0-25 µm, the CF is usually at longer wavelengths, from 7.9-8.1 µm indicating a mineralogical heterogeneity among the finer fraction. Clays are extremely fine-grained and likely remain in the smallest grain size fraction, which could account for the slight shift in the CF (see discussion). The Transparency features (TF) appear at 10.4 - 11.7 µm usually in the smallest size fractions of the bulk suevites. The strongest feature in this region is between 11.6 and 11.7µm. The strong Reststrahlen band (RB) at 9.4 µm again indicates a highly amorphous and homogeneous internal structure of the minerals comprising the sample. Minor bands or shoulders are found on the slope of the RB at 8.5-8.6 µm and 8.8-8.9 µm, indicating crystalline species (see Discussion). Further RB or potential TF are observed in the bulk suevites at 10.4-10.5 µm and 11.1-11.2 µm (overlapping with the transparency feature in the smallest size fractions) and 12.5-12.9 µm. Broad features in the 18-19 µm region are also typical. All bulk suevite spectra show very strong water features at 2.8 µm and 6.1 µm (Fig. 4d). These volatile bands probably result from clay minerals, but analysis under atmospheric conditions could also have affected the spectra. A spectrum of separated fine-grained matrix from Aumühle 18 (Fig. 4a, Tab. 2a) is very similar in band shape and peak positions to the bulk spectrum Aumühle 18, indicating a high abundance of melt and crystalline clasts in the fine-grained matrix.

Melt glasses from Aumühle and Otting (Fig. 4b; Tab. 2b) are also very homogeneous. The Christiansen Features (CF) are at 7.6-7.9 µm, the Transparency Features (TF) at 11.7-11.8 µm in the finest size fraction. The strongest Reststrahlen band is at 9.3-9.4 µm, with a weak shoulder in the finest size fraction at 8.6 µm. This all confirms the amorphous nature of the material. Only weak bands are found in the 18-19 µm region. Water features occur at 6.1 µm and 2.8-2.9 µm (Fig. 4d).

In the red suevite (Polsingen) (Fig. 4c; Tab. 2c), the CF occurs between 7.6 and 7.9 µm. A strong TF band is located at 11.8-12.0 µm, and the strongest Reststrahlen band is at 9.4-9.5 µm. Further, weak Reststrahlen bands or shoulders are between 8.2 and 8.8 µm in red suevite samples. Also, the slopes between 10 and 12 µm in Polsingen 3 and 4 are less steep than in the normal suevite. Broad bands at ~17 µm, and between 18-19 µm are also typical. All red suevites show strong water features at 6.1 and 2.7-3.0 µm, which are probably caused by weathering phases (Fig.4d).

**3.4 In Situ Analyses**

Fine-grained matrix analyses for Aumühle (Fig.5, Table 3) show CFs between 7.6 and 8.3 µm, those for Otting are located at 7.6-8 µm. The only powder analysis of separated matrix from Aumühle 18 overlaps with these ranges (7.4-7.9µm). The strongest Reststrahlen band is at 9.5-9.6 µm in the Aumühle matrix, at slightly longer wavelengths than in the powdered sample (9.4 µm). In the fine-grained Otting matrix, the Reststrahlen band is very similar (between 9.4 and 9.5 µm) for both in situ and powder measurements. There are several additional Reststrahlen bands for fine grained matrix in the in situ analyses: at 8.5-8.6 µm, 8.9-9.0 µm, and from 12.5 to 12.9 µm. The intensities vary from shoulders to clear bands. Here the band positions are very similar to the powdered matrix from Aumühle 18.

The in-situ analyses of glasses in the polished sections of Aumühle and Otting (Fig.5, Tab.3) have Christiansen Features between 7.7 and 7.9 µm, identical to the powder measurements. The dominating Reststrahlen bands are at 9.4 to 9.6 µm, thus slightly shifted to longer wavelengths compared to the powder data. The Polsingen samples where the 'matrix' consists of melt glass, show a high homogeneity: the CF is at 7.9-8 µm, the Reststrahlen band at 9.6 µm. Characteristic is a shoulder/feature at 8.8 µm. Again, a slight shift for the position of the Reststrahlen band was observed compared to the powder analyses, whereas the other features occur at similar wavelength positions.

## 4. Discussion

The investigated impact melt glasses in Aumühle and Otting do not have significant bands of crystalline features, and are dominated by the Reststrahlen band at 9.3-9.4 µm in the powdered samples and 9.4-9.6 µm in the micro-FTIR analyses. This range is similar to quartzofeldspathic glasses or glass with granitic/rhyolitic composition (9.1-9.8 µm) with a mafic component (9.4-10.5 µm) (Wyatt et al., 2001; Byrnes et al, 2007; Johnson et al., 2007; DuFresne et al., 2009; Lee et al., 2010; Minitti and Hamilton, 2010; Wright et al., 2011). This mixture reflects the starting composition of the Ries impact site, which was dominated by gneiss, granite and about 13% mafic rock (amphibolite; von Engelhardt, 1997). The CF of the melt glasses (7.6-7.9 µm) also falls into the region for acidic and intermediate rocks, while mafic minerals and rocks tend to have the CF greater than 8 µm (Pieters and Englert, 1993; Salisbury and Walter, 1989; Cooper et al., 2002).

Bulk melt rock, i.e. the red suevite from Polsingen, shows the main RB band slightly shifted by about 0.1 µm to longer wavelengths. The similarity of the various samples even on millimeter scale in in-situ analyses confirms that the red suevite is a homogeneous melt rock. Compared to the melt glasses in normal suevite, the spectra from Polsingen show weak, but clear RB or shoulders of crystalline materials. These indicate higher contents of crystalline components from granite and gneiss fragments (e.g., Hecker et al., 2012).

Bulk suevites also form a homogeneous group, only the relative intensity of the mostly weak crystalline bands hints at varying contents of partially shocked fragments. Separated matrix is spectrally similar to the bulk material, indicating a high amount of fine-grained rock fragments in the matrix. This is confirmed by the in-situ analyses of the fine-grained matrix, which have the same/similar band positions, but varying intensities of the Restrahlen bands. The Reststrahlen bands in bulk powder suevite and fine-grained matrix are a mixture of features characteristic for glasses and crystalline components.

The dominating band in the bulk samples between 9.4 µm (powders) and 9.4 to 9.6 µm (micro-FTIR) is similar to the melt glasses (Byrnes et al, 2007; Lee et al., 2010; Wright et al., 2011). However, crystalline features between 8.5-8.6 µm, 8.8-9.0 µm, and from 12.5 to 12.9 µm in the bulk and fine-grained matrices are identical to that of quartz (e.g., Wenrich and Christensen, 1996; Christensen et al., 2000; Michalski et al., 2003; Baldridge et al., 2009), a major component of biotite-rich granite and gneiss as found in the basement of the Ries Crater (von Engelhardt, 1997; Hecker et al., 2012).

The ranges for the position of the Christiansen Features in glass and bulk suevite overlap, but some of the bulk samples have their CF at significantly shorter wavelengths compared to the glass. This probably reflects the increased content of a crystalline quartz component. The lowest CF are between 7.4 and 7.5 µm, close to that of quartz with 7.35-7.4 µm (Tab. 2a)(e.g., Christensen et al., 2000; Michalski et al., 2003; Baldridge et al., 2009; Tappert et al., 2013). This in turn would point to a significant crystalline component at least in some of the bulk suevites. A comparison with CF positions of terrestrial rocks (Salisbury and Walter, 1989; Cooper et al., 2002) also indicates that the bulk powder suevites (7.4-8.1µm) cover the range for acidic rocks (7.6-7.8 µm) and intermediate rocks (7.8-8.2 µm).

The variations in the positions of the CF are probably due to slight compositional variations as a result of sieving. A higher content of quartz drives the CF to shorter wavelengths (e.g., Christensen et al., 2000; Michalski et al., 2003; Baldridge et al., 2009; Tappert et al., 2013), while an increase of clay minerals or a mafic component (amphibolite) in the finest fraction during sieving could shift the position of the CF towards longer wavelengths. Hornblende, a major phase of amphibolite, which comprises the mafic component in the suevite (von Engelhardt, 1997), has a high CF from 8.2-8.5 µm (Salisbury, 1992; 1993; Baldridge et al., 2009). The CF of alteration phase montmorillionite falls between 7.9 and 8.2 µm and would be difficult to identify directly in a mixture with rocks of intermediate composition (Christiansen et al., 2000; Cooper et al., 2002; Koeppen et al., 2005; Michalski et al., 2005; Baldridge, 2009).The CF shift is also pronounced in the micro-FTIR spectra of the Aumühle samples, where the intensity of the crystalline quartz features also varied in the analyzed areas (Fig.5).

In summary, in the mid-infrared range, suevites can be identified by their highly amorphous band shape and few crystalline features. This could help to identify impactites like suevite in remote sensing data, but also provides information about the shocked basement rock. However, glassy, amorphous spectral features can also be expected from volcanic glasses like obsidian, so additional geologic information about an observed area is necessary to distinguish the sources of glassy material (Hamilton et al., 2001; Moroz et al., 2009; Wright et al., 2011).

While the starting composition of Mercury was probably not like the granitic basement of the impact site in the Ries, i.e., in terms of mineralogy, it is still possible to draw conclusions for future observations of the Mercurian surface. Mercury is a planet that underwent massive impact cratering in its early history (e.g., Hiesinger et al. 2010; Strom et al., 2011; Fassett et al., 2012). As a consequence, the resulting future mid-infrared observations that will be made of Mercury by MERTIS on BepiColombo might show only a few strong crystalline features. While the Ries crater represents only the effects of one impact, surface material on Mercury underwent many impacts resulting in regolith gardening (Domingue et al., 2014). The low degree of crystallinity of all involved rocks in suevite after only one impact event indicates that the degree of amorphization can be expected to be much higher on the surface of Mercury.

It has to be taken into account that the suevite is not the only type of ejecta observed in the Ries. By volume, it is only a relatively small component – the Bunte Breccia has an estimated volume of 95 km$^3$, the megablocks about 47 km$^3$, and up to 22 km$^3$ of suevite (Stöffler et al., 2013; Sturm et al, 2015). However, the suevite is the uppermost ejecta layer (e.g., von Engelhardt, 1990), and so is the material most likely to be observed in remote sensing. Also, in a continuing impact gardening of a planetary surface, the less shocked rocks will also increasingly experience higher degrees of shock metamorphism. This might be even more enhanced by higher impact velocities on Mercury compared to Earth, which result in the production of larger amounts of impact melt (e.g., Fassett et al., 2012).

In addition, the effects of space weathering will damage the crystalline structure of the remaining material even more (Domingue et al., 2014). Observations in the ultraviolet, visible and near-infrared range by MESSENGER do not show much variation over the surface, which could also point towards a complete amorphization of the surface minerals (Izenberg et al., 2014). However, constant reprocessing and gardening of the surface may also allow formation of new crystallites, resulting in a mixture of crystallized and amorphous material. In addition volcanic activity (Thomas et al., 2014) will produce crystalline phases. Consequently, ground based spectroscopic observations of larger areas on Mercury (e.g., Sprague et al., 2000) show features of crystalline species such as pyroxene and feldspar.

Furthermore, for our studied samples, alteration of pristine impact material by weathering has to be taken into account. The fine-grained parts of the matrix contain clay minerals, mainly montmorillonite, but also minor illite and halloysite (Stöffler et al, 2013). A strong band of montmorillonite is found at 9.4-9.5 µm, overlapping with the position for the strong feature of amorphous silicates. The strong feature of illite and halloysite is at slightly longer wavelengths, 9.4-9.7 µm (Christensen et al., 2000; Koeppen et al., 2005; Michalski et al., 2006; Baldridge et al., 2009). A further feature that could indicate montmorillonite is a weak band at about 8.8 µm (Christensen et al., 2000; Michalski et al., 2006; Baldridge et al., 2009), occurring in most bulk suevites and the Polsingen sample (Tab.2). However, this band is also characteristic for quartz (e.g., Christensen et al., 2000; Michalski et al., 2003; Tappert et al., 2013).

Another way to identify clay minerals are spectral features at longer wavelengths, in the 17-19 µm regions. Clay minerals have features in the 18-19 µm region, montmorillionite from 18.7-19 µm, illite and halloysite from 18-19 µm (Christensen et al., 2000; Baldridge et al., 2006; Michalski et al., 2006). Here the Aumühle and Otting glasses show only a few, weak features. The suevite bulk samples show more features in the 18.0-19 µm region, especially from 18.4-19 µm. This points towards a clay component, but quartz can also have a feature at 18.3 µm (Baldridge et al., 2009). In contrast to the Aumühle and Otting glasses, the Polsingen melt rocks show clear bands in the 18-19 µm region, indicating a stronger degree of weathering compared to the glasses. However, this part is at the limit of the spectral range of the spectrometer used, so exact band positions are difficult to obtain.

The strongest transparency feature of the bulk suevites (11.6-11.7µm) and red suevite (12.0µm) is similar to that of the Aumühle and Otting glasses (at between 11.7 – 11.8 µm), and is also typical for acidic and intermediate rocks (Salisbury and Walter, 1989; Cooper et al., 2002). A potential TF of montmorillionite and illite at 12.1-12.7 µm is not visible in the smallest size fractions, except possibly the Red Suevite from Polsingen. Illite has another TF at 11.5-11.7, which would overlap with the glass TF (Baldridge et al., 2006). A potential TF in bulk suevite at 11.1-11.2 µm is similar to one of the montmorillonite TF found at 11.2-11.5 µm (Baldridge et al., 2006). However, quartz has its TF also in this region (10.9-11.1µm) (Salisbury 1992; 1993; Baldridge et al., 2009).

Water bands could also potentially allow identifying secondary phases, but analyses were conducted under ambient air pressure. The water bands for the samples in this study are all very similar to each other, which indicate that adsorbed water could have influenced the spectra. This would render the features difficult to use for comparison.

Further hints for the occurrence of smectites are elevated aluminum contents in the matrices and the Polsingen samples (Tab.1). Certainly, aqueous alteration due to volatiles appears implausible on Mercury at large scales. But application of the data from this study for remote sensing observations of bodies where impactites were affected by alteration, like Earth or Mars, is still valid, as (minor to moderate) alteration features do not overprint the spectral signature of the impactites.

5. Conclusion

Despite the general similarity of their bulk spectra, bulk suevite, red suevite melt rock, and impact melts are distinguishable by their mid-infrared features, although they are all dominated by amorphous materials. Suevite glass from Aumühle and Otting shows simple spectra, dominated by an amorphous feature. Suevites have clear bands of crystalline materials, while the Polsingen impact melt falls between the two groups.

Secondary phases like clay minerals have features overlapping with other components (lithic rock clasts, amorphous material) and are difficult to identify in the laboratory bulk spectra. This shows that low to moderate amounts of alteration may not significantly affect the study of impactites on remote bodies.

The samples of Polsingen impact melt also show signs of weathering, in contrast to the melt glasses from Aumühle and Otting.

On the basis of our observations, we conclude that in mid-infrared remote sensing data, the surface layer of Mercury will be dominated by features of amorphous materials. Because of the high degree of amorphization that occurs after only one impact, as inferred from the Ries event, and following impact gardening, even remaining crystalline materials will undergo high degrees of shock metamorphism. However, the suevite represents only the uppermost impact layer in the Ries, while the underlying, more voluminous ejecta show higher degrees of crystallinity.

**Acknowledgements**

Many thanks to Prof. Alexander Deutsch (Münster) and Gisela Pösges (Rieskrater Museum, Nördlingen) for precious help with the samples. We also thank the editor, Will M. Grundy, Steve Ruff and another anonymous reviewer for helping to improve the manuscript.

This work is supported by the DLR funding 50 QW 1302 in the framework of the BepiColombo mission.

**Figure Captions**

**Figure 1:** Map of the Ries area (adapted from Stöffler, 2013). Samples for this study were taken in the Otting, Aumühle and Polsingen locations.

**Figure 2a-c:** Micrographs of representative areas of locations sampled for this study under normal transmitted light (top) and crossed polarizers (bottom). The Otting (a) and Aumühle (b) samples show the melt glasses embedded in the fine grained matrix. Spots (1 mm$^2$ each) analyzed in situ with a FTIR-microscope are marked with black boxes. Red suevite from Polsingen (c) does not clearly show the distinction between amorphous material and matrix, it is probably a coherent melt rock.

**Figure 3:** Comparison of SEM/EDAX data for melt glass and matrix. Data from this study: squares = Aumühle, circles = Otting, triangles = Polsingen. Empty symbols = glass; filled symbols = fine-grained matrix. Literature data (Stöffler et al., 2013) are marked with encircled areas (see legend). Chemical data for impact melt glass in suevite from our study fall into an area for earlier analyses. Analyses of fine-grained matrix overlap with bulk–suevite data, but also fall outside the range, due to higher contents of secondary phases.

**Figure 4a-d.** Mid-infrared reflectance spectra of samples in the size fractions 0-25µm (blue), 25-63µm (pink), 63-125µm (red) and 125-250µm (brown) of (a) Suevite bulk and matrix analyses from Aumühle and Otting, (b) Spectra of separated melt glasses from Aumühle and Otting, (c) Samples of red suevite melt glass from Polsingen. Vertical lines mark characteristic Christiansen, Reststrahlen and Transparency features. (d) Water features in the range from 2-7 µm for selected samples. Spectra are off-set for clarity. Blue: melt glasses, Red: red suevite from Polsingen, Purple: suevite bulk samples.

**Figure 5:** Infrared and Raman for Interplanetary Spectroscopy (IRIS) laboratory In situ mid-infrared spectra obtained from polished sections of suevite using a FTIR microscope. Each area was 1 mm$^2$ in size; the locations are shown in Figure2 a-c.

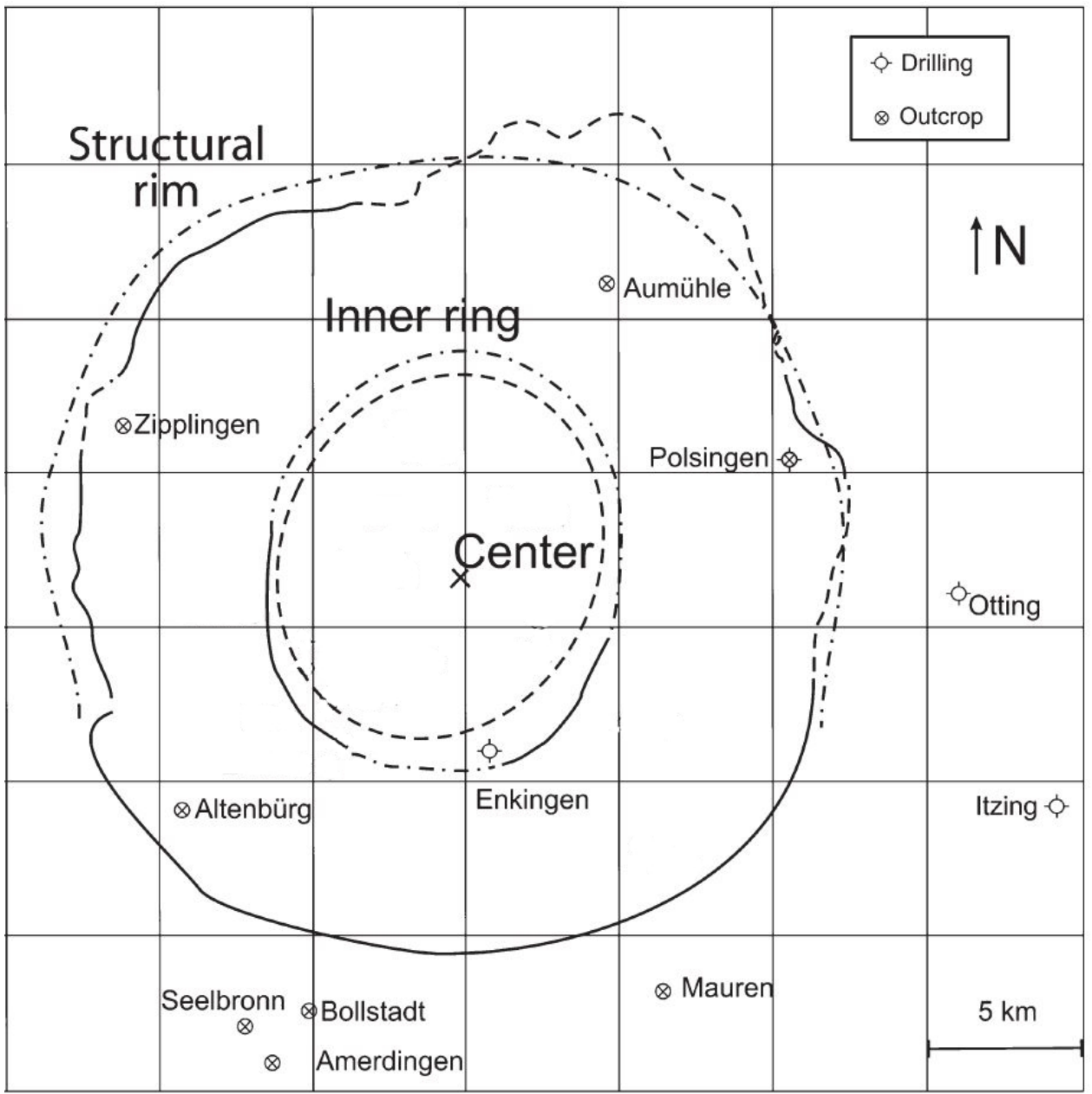

Figure 1

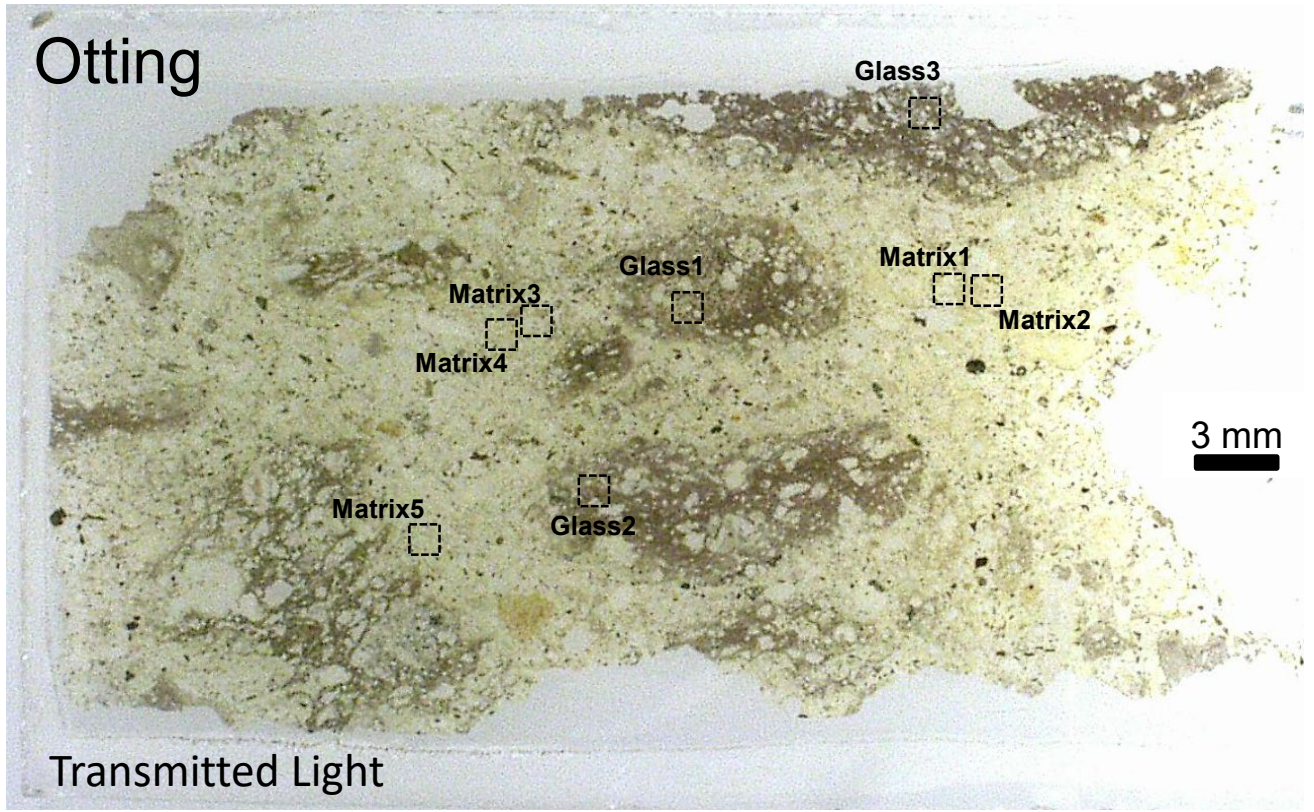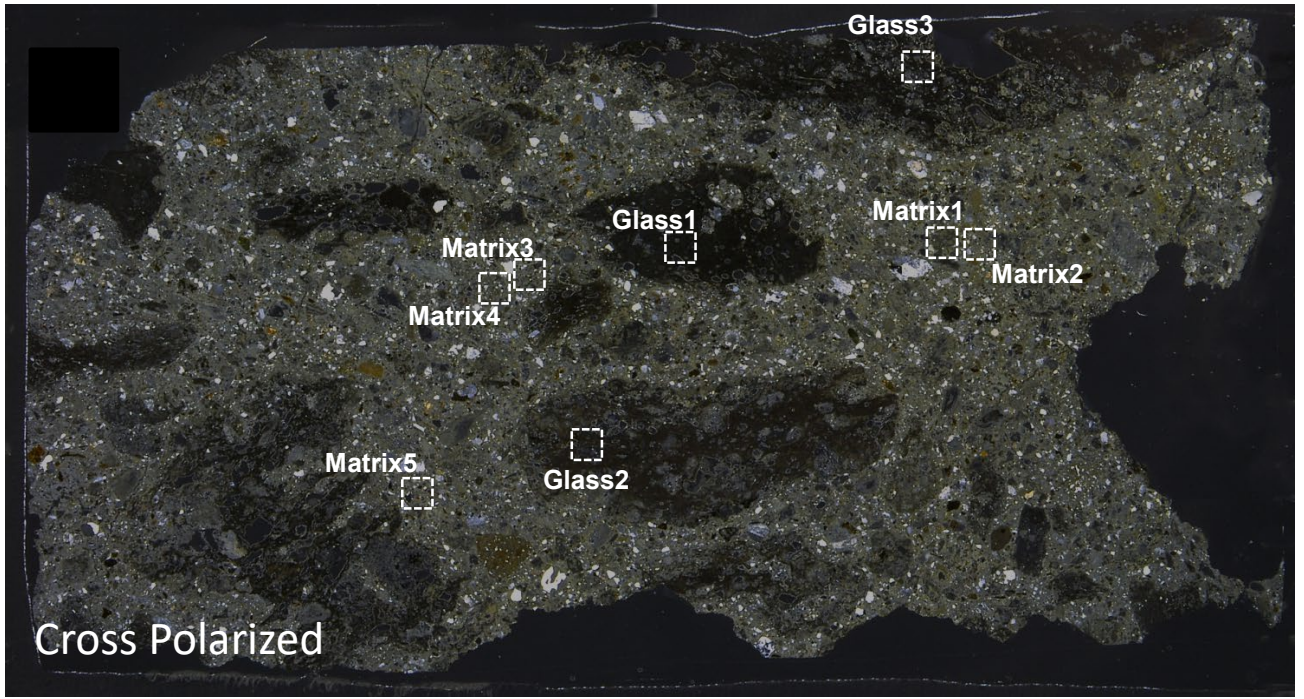

Figure 2a

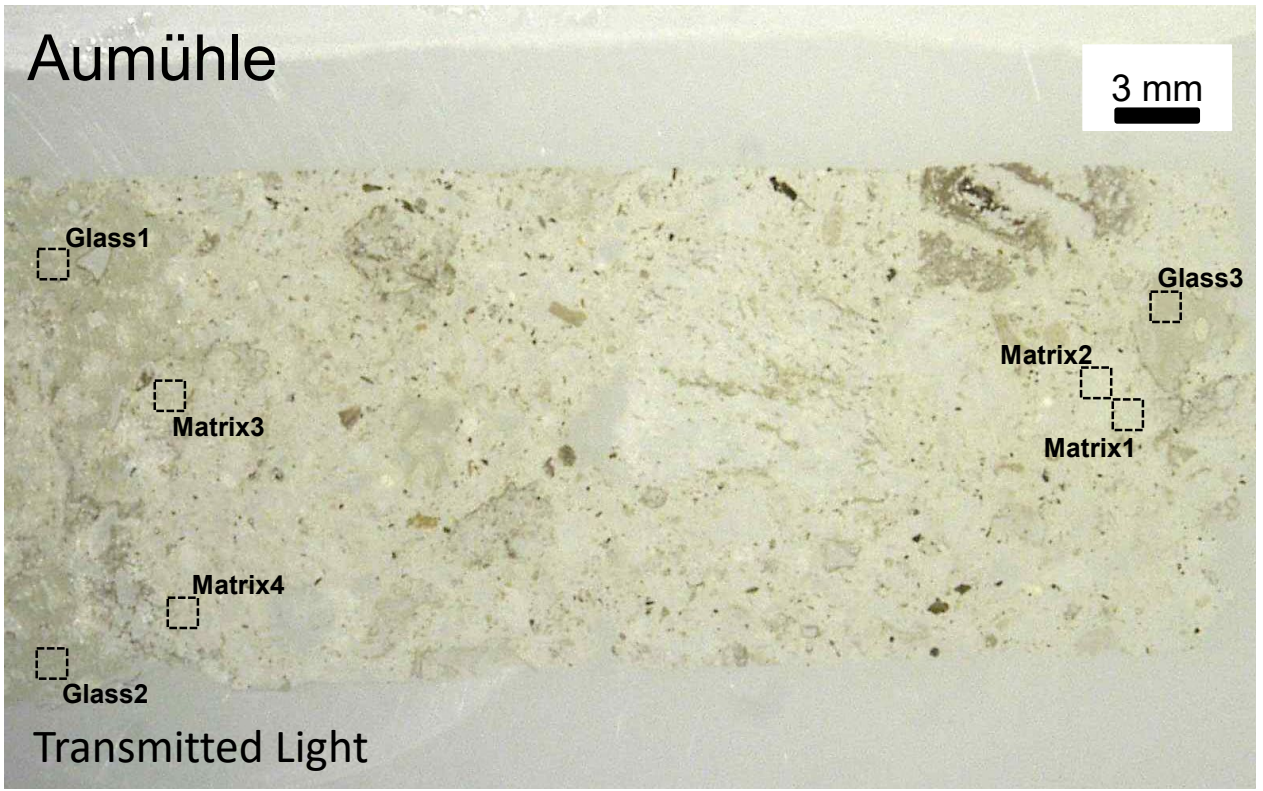
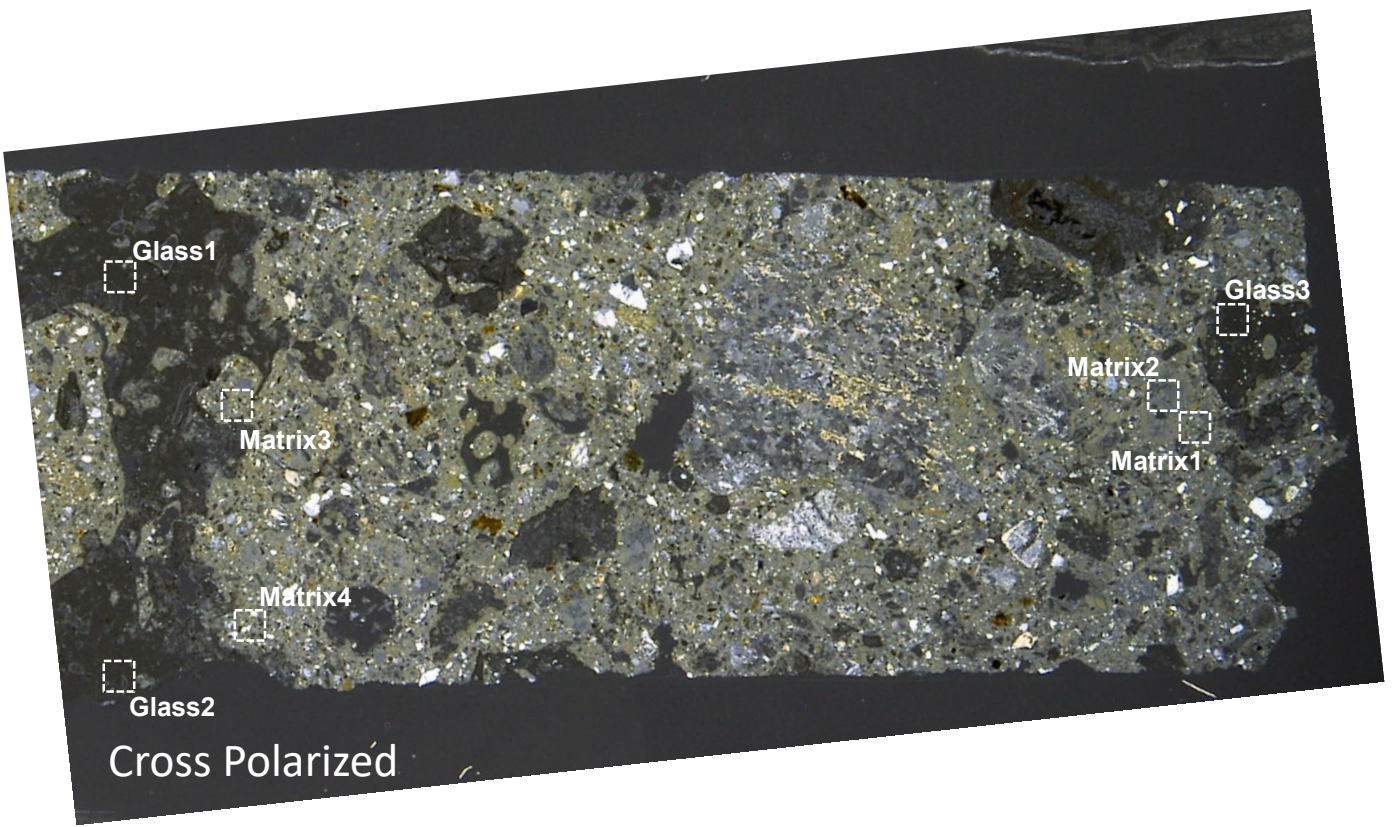

Figure 2b

# Polsingen

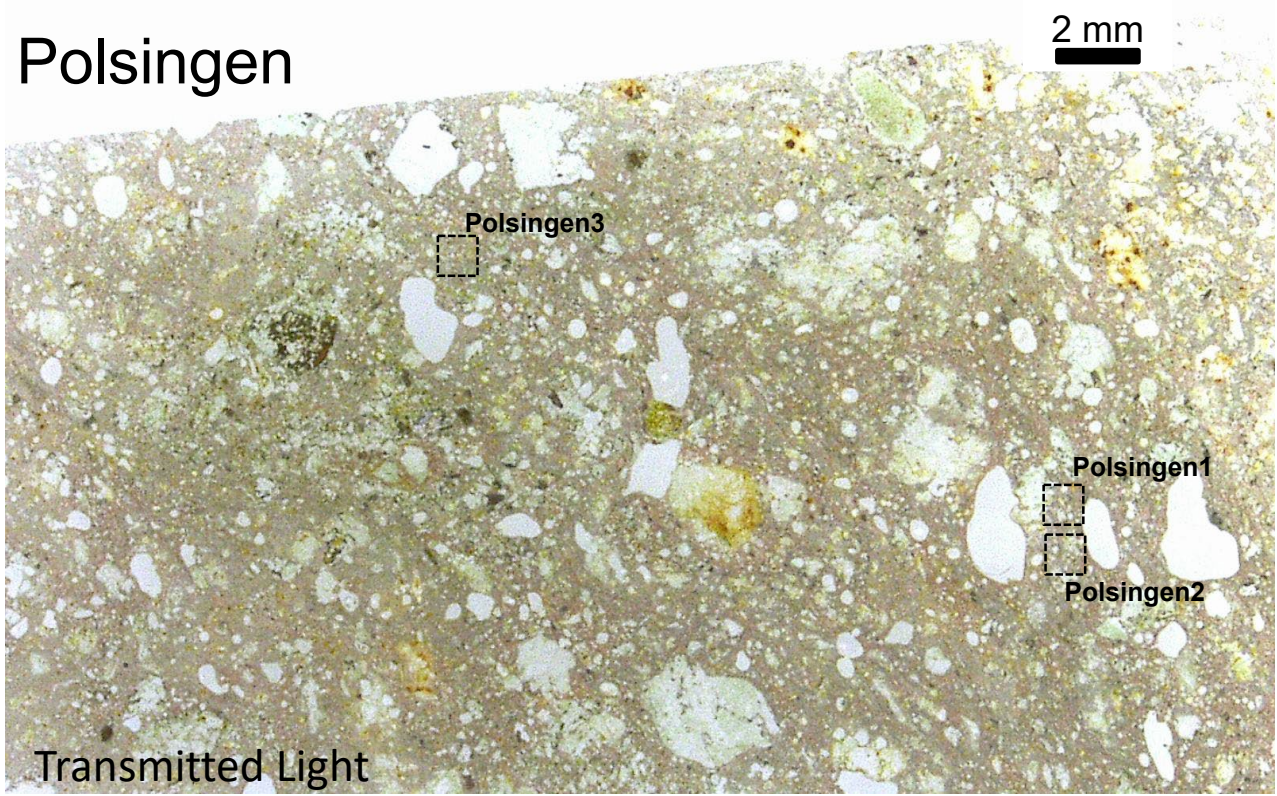

Transmitted Light

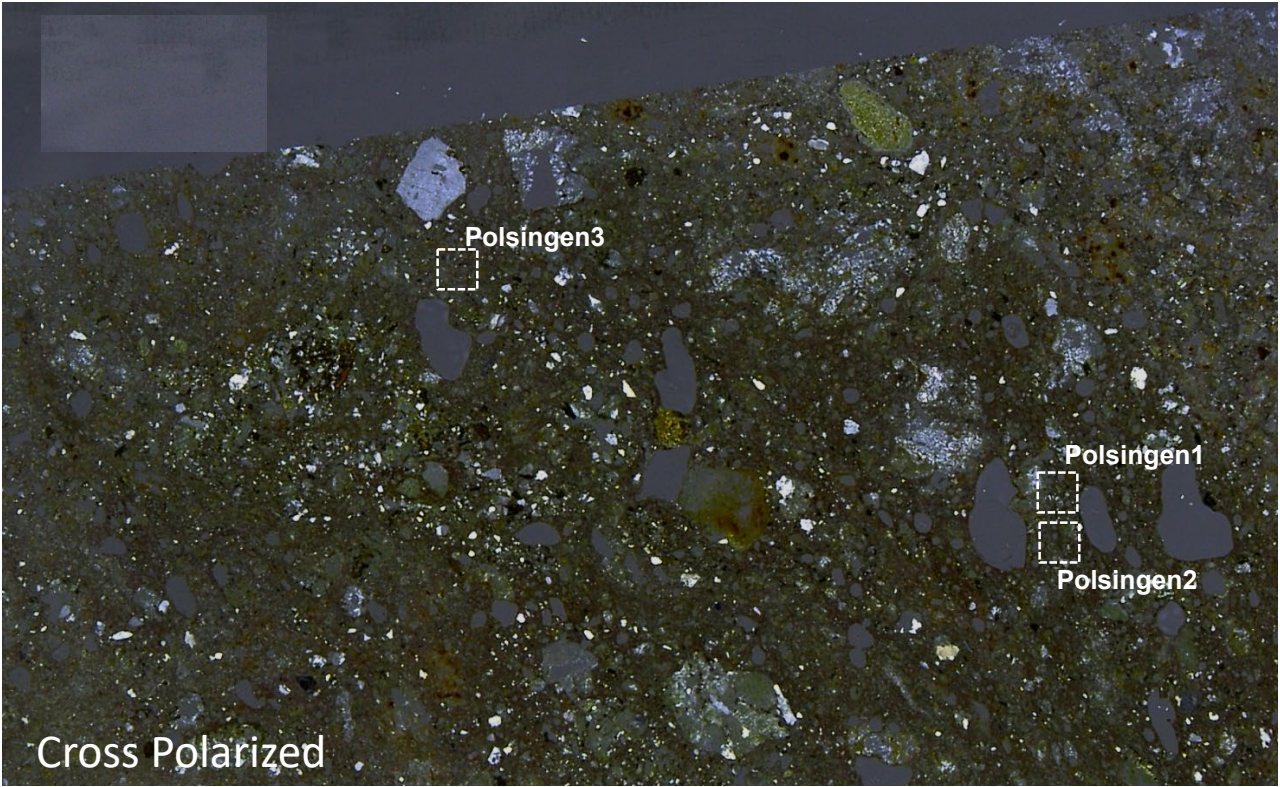

Cross Polarized

Figure 2c

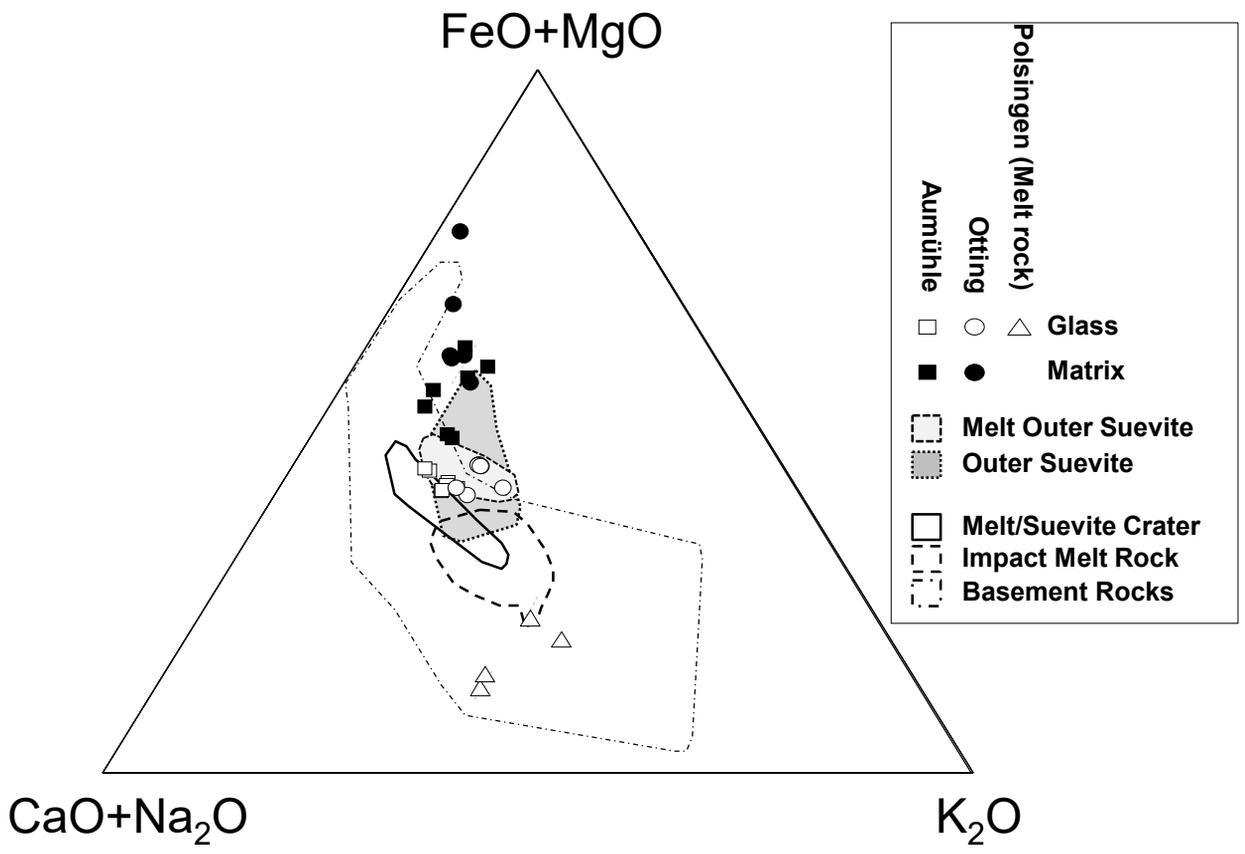

Figure 3

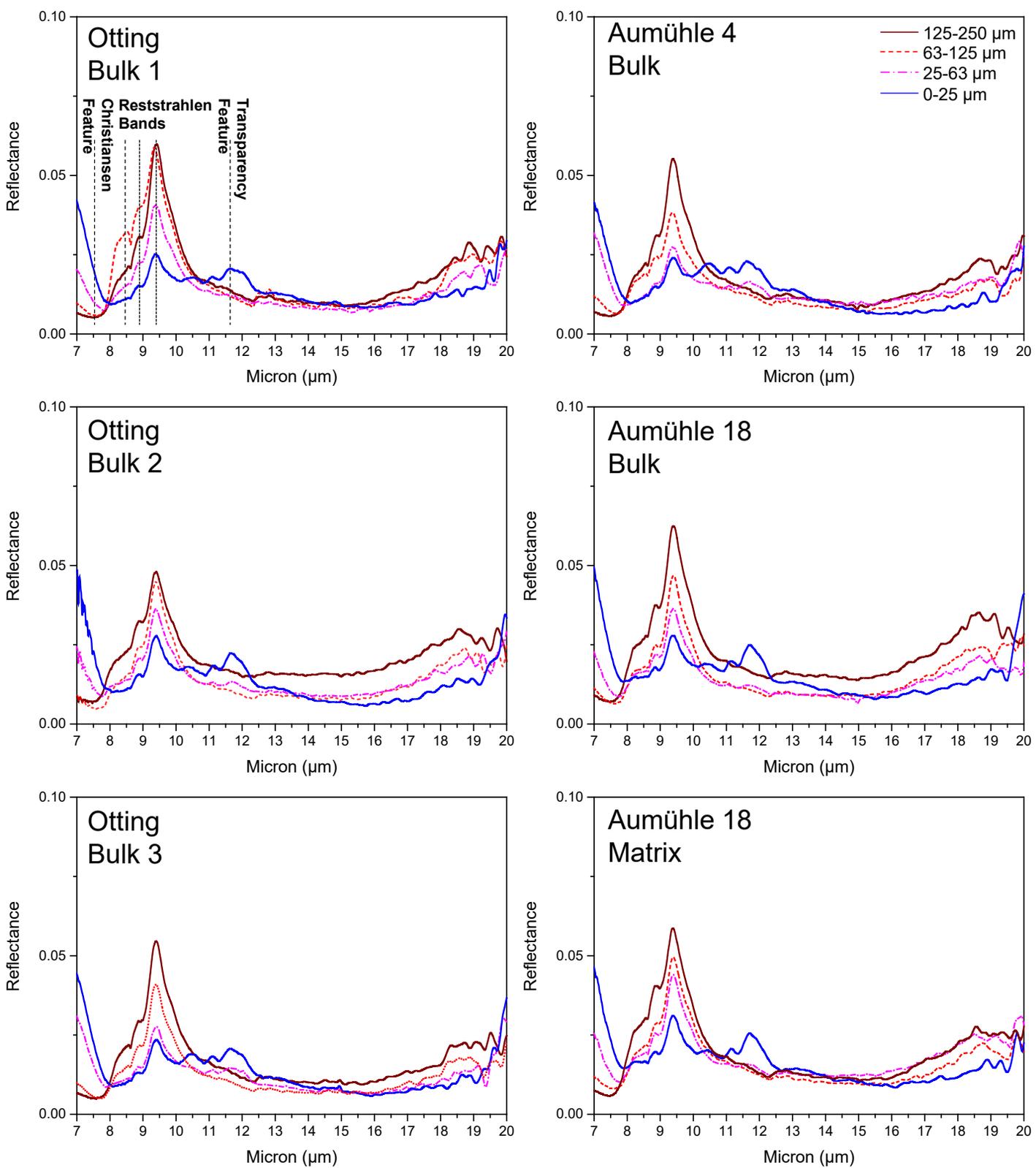

Figure 4a

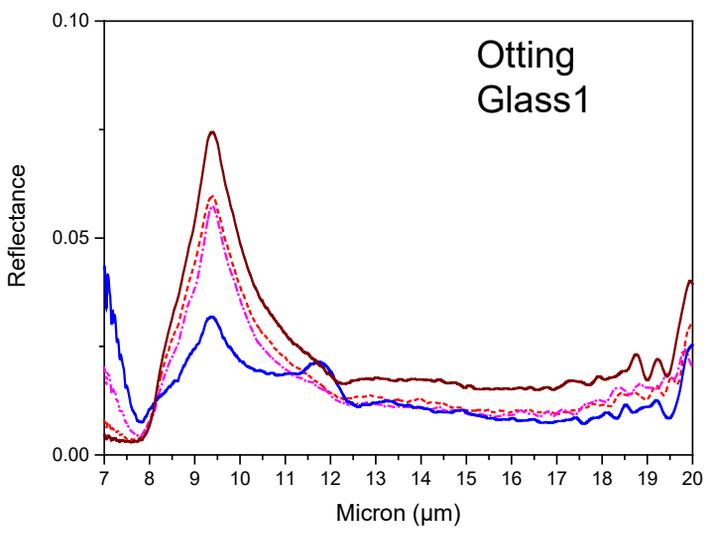
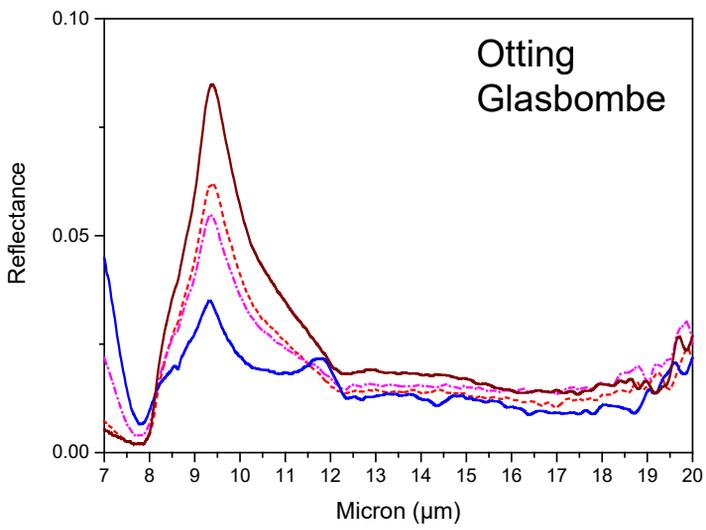
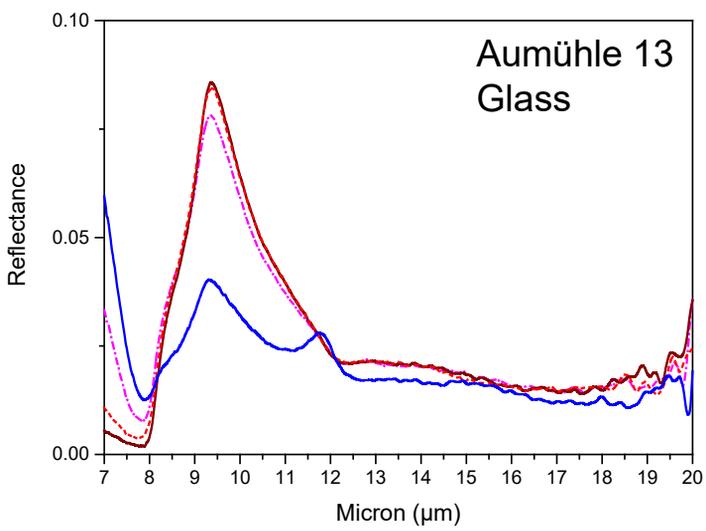

Figure 4b

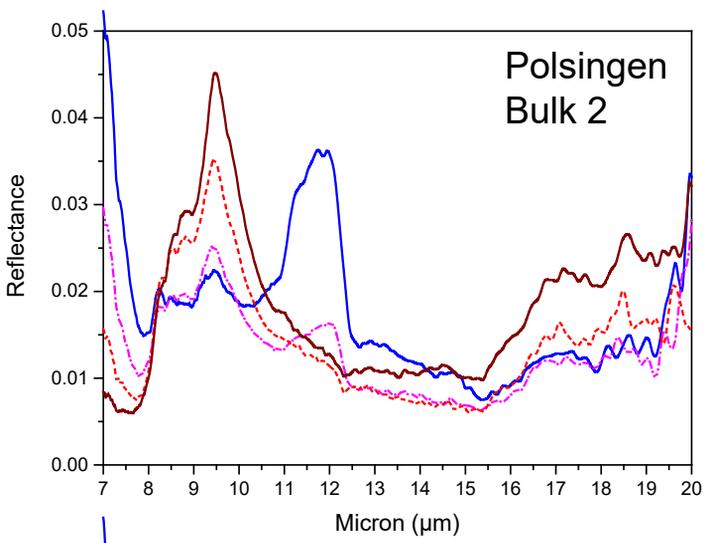
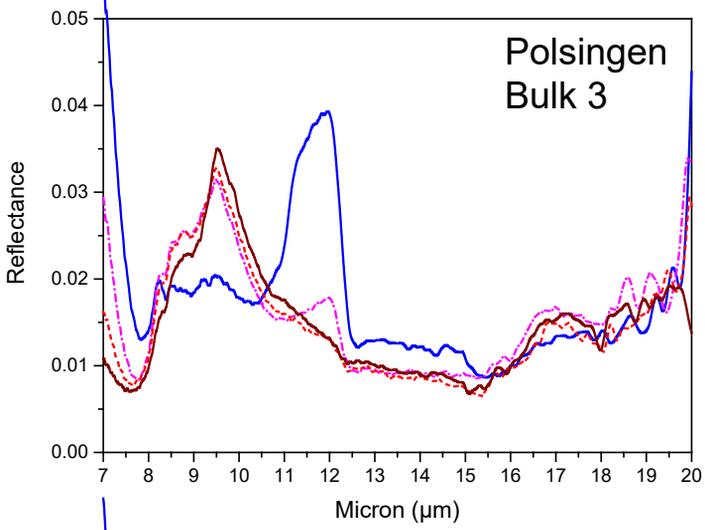
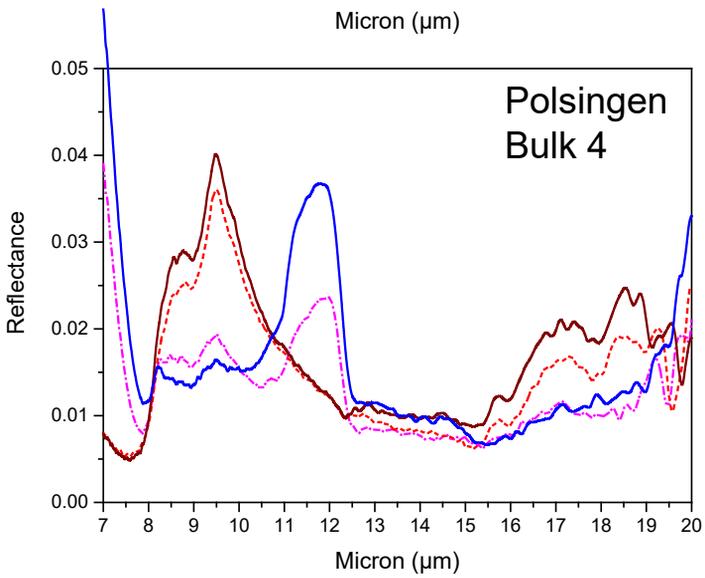

Figure 4c

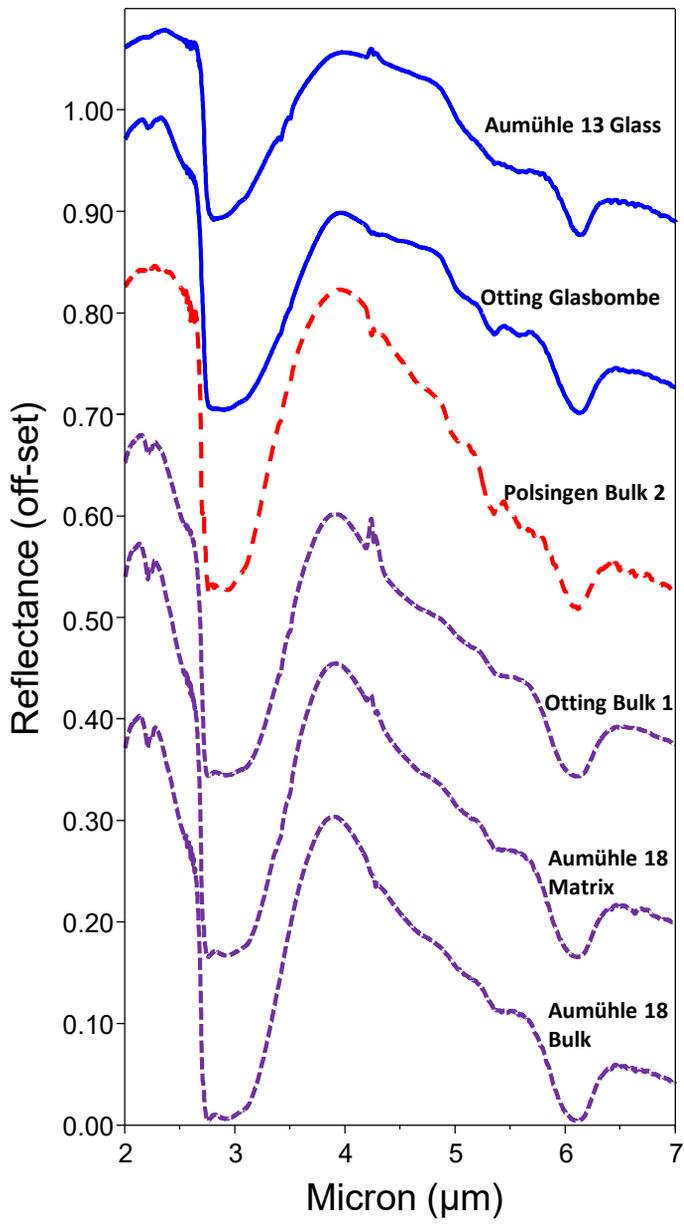

Figure 4d

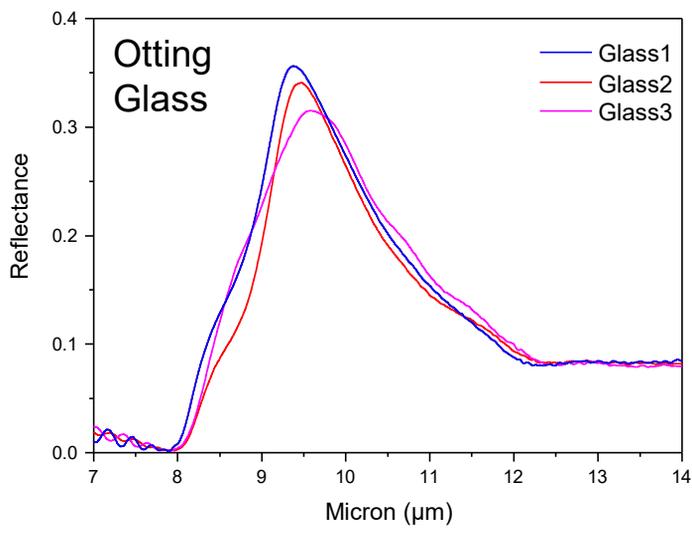
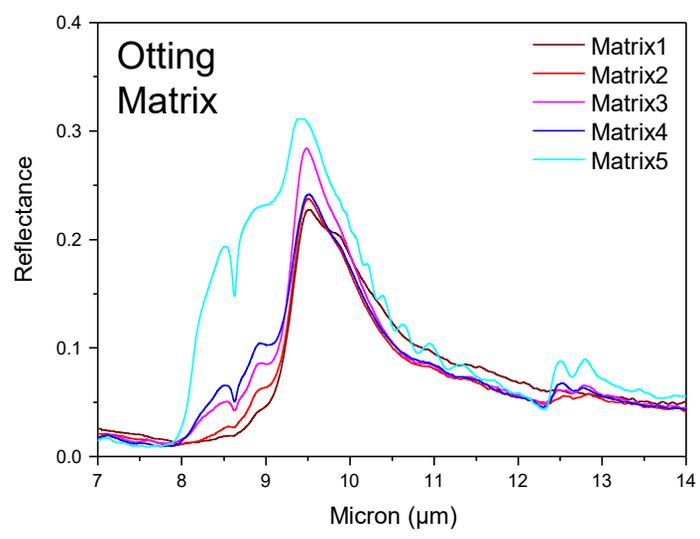
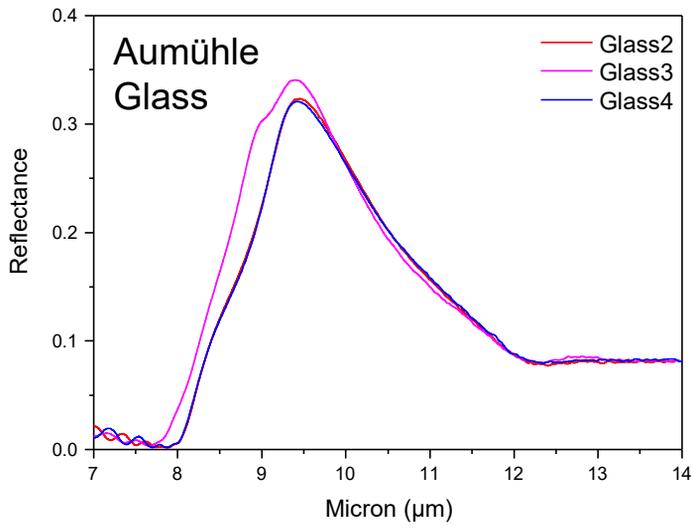
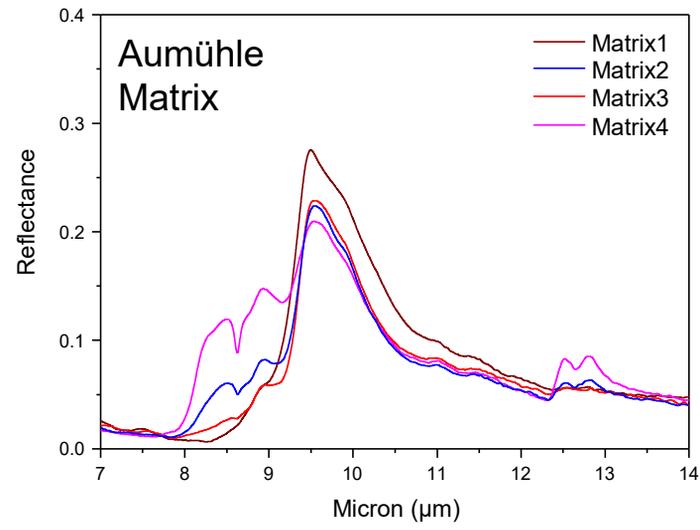
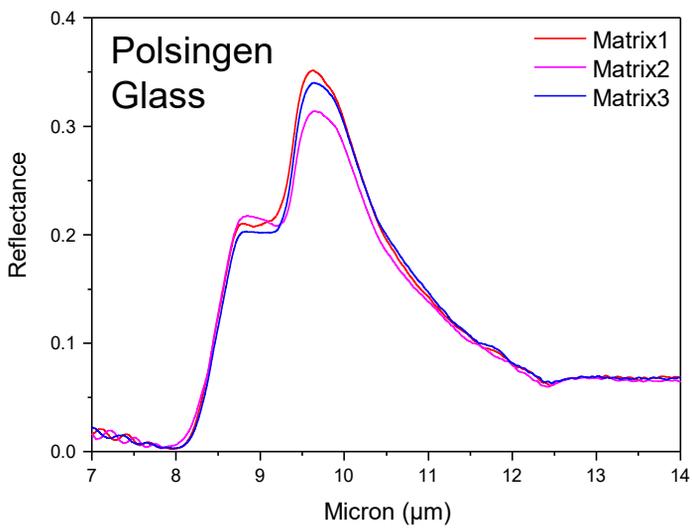

Figure 5